\documentclass[final,5p,twocolumn,sort&compress,number]{elsarticle}
\usepackage{amssymb}
\usepackage{amsmath}
\usepackage{graphicx}

 \def\bc{\begin{center}}          \def\ec{\end{center}}
 \allowdisplaybreaks

\begin{document}
\title{WARP-X: A NEW EXASCALE COMPUTING PLATFORM FOR BEAM-PLASMA SIMULATIONS}

\author[LBNL]{J.-L. Vay}\ead{jlvay@lbl.gov}
\author[LBNL]{A. Almgren}
\author[LBNL]{J. Bell}
\author[SLAC]{L. Ge}
\author[LLNL]{D.~P. Grote}
\author[SLAC]{M. Hogan}
\author[SLAC]{O. Kononenko}
\author[LBNL]{R. Lehe}
\author[LBNL]{\\A. Myers}
\author[SLAC]{C. Ng}
\author[LBNL]{J. Park}
\author[LBNL]{R. Ryne}
\author[LBNL]{O. Shapoval}
\author[LBNL]{M. Th\'evenet}
\author[LBNL]{W. Zhang}

\address[LBNL] {LBNL, Berkeley, CA 94720, USA}
\address[LLNL] {LLNL, Livermore, CA 94550, USA}
\address[SLAC] {SLAC, Menlo Park, CA 94025, USA}

\date{\today}

\begin{abstract}
Turning the current experimental plasma accelerator state-of-the-art from a promising technology into mainstream scientific tools depends critically on high-performance, high-fidelity modeling of complex processes that develop over a wide range of space and time scales. As part of the U.S. Department of Energy's Exascale Computing Project, a team from Lawrence Berkeley National Laboratory, in collaboration with teams from SLAC National Accelerator Laboratory and Lawrence Livermore National Laboratory, is developing a new plasma accelerator simulation tool that will harness the power of future exascale supercomputers for high-performance modeling of plasma accelerators. We present the various components of the codes such as the new Particle-In-Cell Scalable Application Resource (PICSAR) and the redesigned adaptive mesh refinement library AMReX, which are combined with redesigned elements of the Warp code, in the new WarpX software. The code structure, status, early examples of applications and plans are discussed.
\end{abstract}

\maketitle

\section{Introduction}

Particle accelerators are a vital part of the DOE-supported infrastructure of discovery science and university- and private-sector applications, and have a broad range of benefits to industry, security, energy, the environment and medicine. To take full advantage of their societal benefits, however, we need game-changing improvements in the size and cost of accelerators. Plasma-based particle accelerators stand apart in their potential for these improvements. Turning this from a promising technology into mainstream scientific tools depends critically on high-performance, high-fidelity modeling of complex processes that develop over a wide range of space and time scales.

As part of DOE's Exascale Computing Project \cite{DOEECP}, a team from Lawrence Berkeley National Laboratory, in collaboration with teams from SLAC National Accelerator Laboratory and Lawrence Livermore National Laboratory, is developing a new powerful plasma accelerator simulation tool. The new software will harness the power of future exascale supercomputers for the exploration of outstanding questions in the physics of acceleration and transport of particle beams in chains of plasma channels. This will benefit the ultimate goal of compact and affordable high-energy physics colliders, and many spinoff applications of plasma accelerators along the way.

During this project, we are combining three major software components (Warp, AMReX and PICSAR) into a new software tool (WarpX) that will be tuned for running efficiently at scale on exascale supercomputers. Warp \cite{Warp} is a framework for the modeling of plasma, beam and particle accelerators using state-of-the-art parallel Particle-In-Cell (PIC) methods. The Adaptive Mesh Refinement (AMR) library AMReX \cite{AMReX} is a robust implementation of the AMR methodology for various applications. The novel library PICSAR \cite{PICSAR}, developed in collaboration with researchers from CEA Saclay in France, implements the elementary PIC operations with extensive optimizations for many-core architectures.

The code structure, status, early examples of applications and plans are discussed in the following sections.

\section{Code structure description}

\begin{figure}
\center
\includegraphics[width=0.5\textwidth]{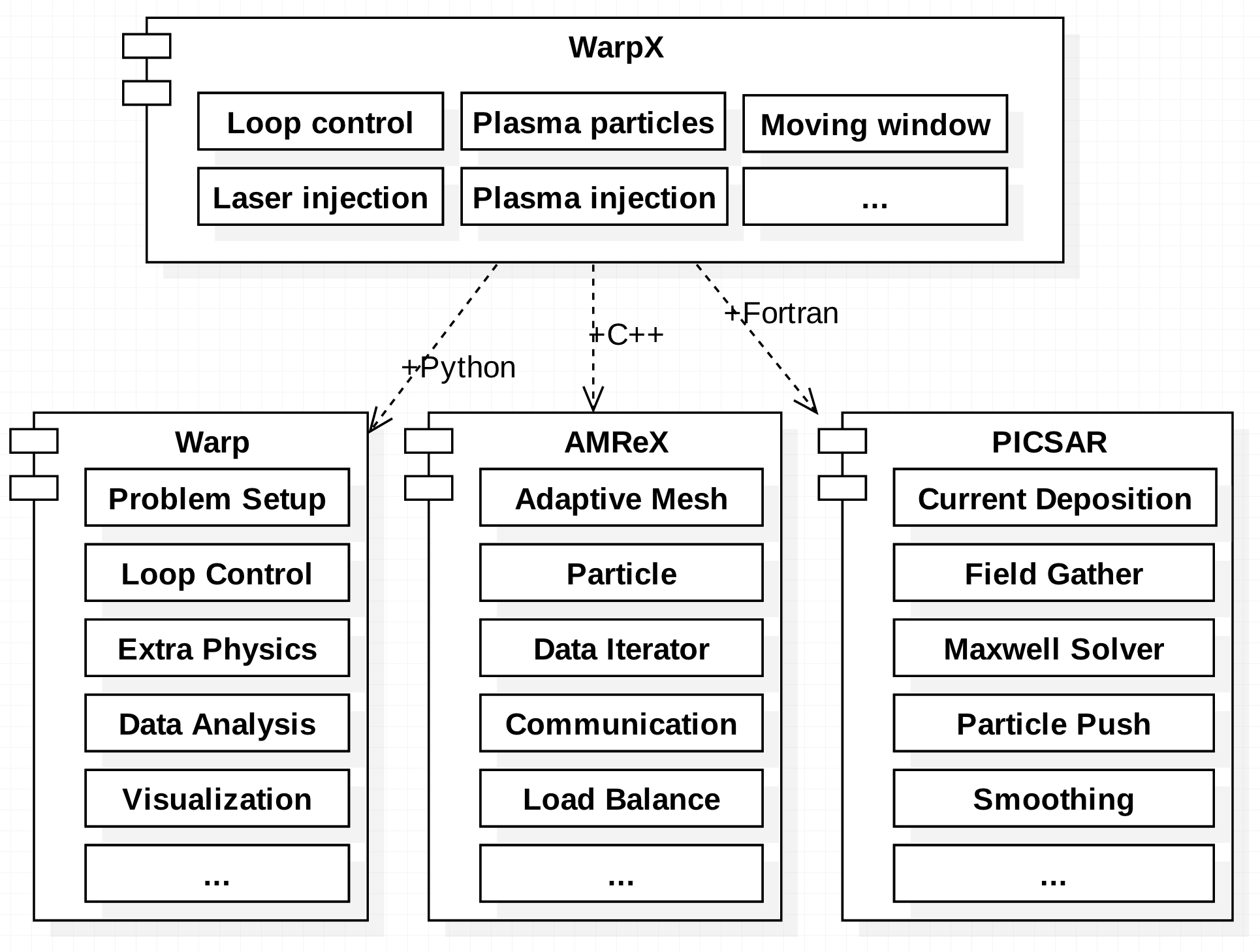}
\caption{\label{fig:WarpXUML} UML diagram of WarpX. In addition to WarpX's own source code (`WarpX-Source'), 
functionalities are provided by three packages: AMReX for the handling
of AMR, communication and load balancing, PICSAR for the low level individual PIC functionalities, and Warp (optional) for extra physics packages, alternate user interface and control. }
\end{figure}

A UML diagram of the WarpX application is given in Fig. \ref{fig:WarpXUML}. The application contains three packages (Warp \cite{Warp}, AMReX \cite{AMReX} and PICSAR \cite{PICSAR}) that are orchestrated together with WarpX's source code (`WarpX-Source'), as follows:

\begin{itemize}
\item Warp \cite{Warp} is a pre-existing PIC code with a Python interface and FORTRAN subroutines for fast number crunching. 
Warp's extensive collection of functionalities can be leveraged by WarpX, when running with Python as the front-end (optional): the Warp framework provides a Python interface with modules for code control, user steering and additional physics (not included in WarpX-Source, AMReX or PICSAR) and diagnostic packages. 

\item The PICSAR library \cite{PICSAR} contains FORTRAN subroutines (originally from Warp) for elemental Particle-In-Cell operations at the innermost loop on grid and particle data, i.e. charge and current deposition from the particles to the grid, Maxwell solver, field gather from the grid to the particles and particle pusher. These subroutines have been (and continue to be) highly optimized for new multicore architectures such as Intel KNL.

\item The PICSAR subroutines are called within intermediate loops for each set of grids and particles of the AMR hierarchy by the AMReX library (C++) \cite{AMReX}, which also handles intra-node (OpenMP) and inter-node (MPI) parallelism, load balancing and parallel I/O on those data. 

\item WarpX-Source is a set of C++, FORTRAN and Python subroutines for interfacing Warp, AMReX and PICSAR. WarpX-Source does not just stitch together the three other components, but orchestrates the workflow at the core of the main loop level. WarpX-Source contains an alternate C++ driver for convenience in development, testing, profiling, and running on systems where Python support is problematic. It also contatins some recoded pre-existing Warp functionalities, as needed for compatibility with AMReX.

\end{itemize}

\section{Status}
\begin{figure*}[h]
\begin{centering}
\includegraphics[width=0.9\textwidth]{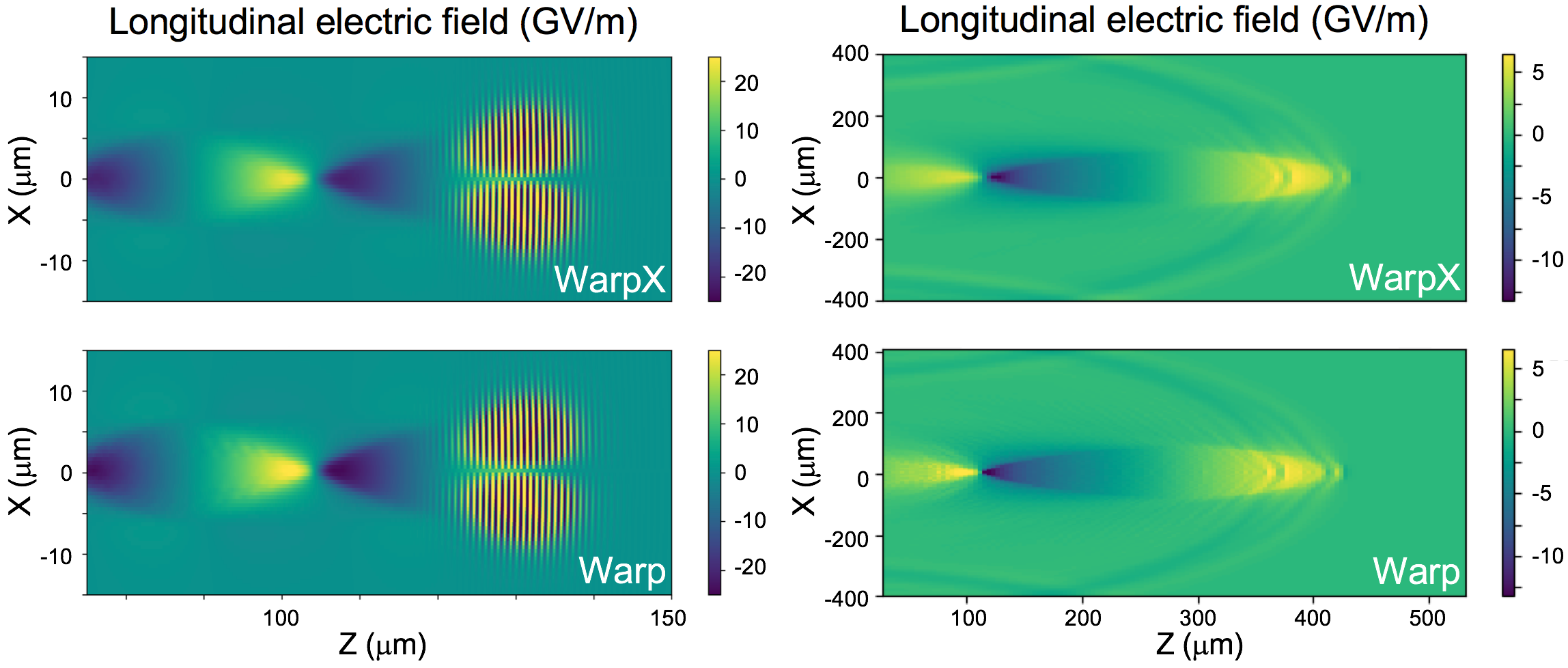}
\par\end{centering}
\caption{Electric $E_z$ field in a laser-driven (left) and particle-driven (right) plasma accelerator simulations using WarpX (top) or Warp (bottom)}\label{fig:pba_lpa_comp}
\end{figure*}

During the first year of development, the three pre-existing components (Warp, AMReX, PICSAR) were combined, and the new code was benchmarked against Warp, first on simple test problems (e.g. Langmuir oscillations, relativistic plasma drift inducing numerical Cherenkov, etc), then on the modeling of particle beam-driven plasma accelerator (BPA) and laser-driven plasma accelerator (LPA). Fig.~\ref{fig:pba_lpa_comp} shows a slice of the longitudinal electric field $E_z$ in the $y=0$ plane, from WarpX and Warp 3-D simulations. The driver beam (laser or electrons) propagates towards the right and drives a wake with strong electric fields. WarpX and Warp results are in excellent agreement.

Mesh refinement was then implemented in WarpX, following the method that had been implemented and validated in Warp \cite{VayCSD12}, based on the following principles: i) avoidance of spurious effects from mesh refinement, or minimization of such effects; ii) user controllability of the spurious effects' relative magnitude; iii) simplicity of implementation. The two main generic issues that were identified in earlier work are: a) spurious self-force on macroparticles close to the mesh refinement interface \cite{Vaylpb2002,Colellajcp2010}; b) reflection (and possible amplification) of short wavelength electromagnetic waves at the mesh refinement interface \cite{Vayjcp01}. The two effects are due to the loss of translation invariance introduced by the asymmetry of the grid on each side of the mesh refinement interface. The implementation that was adopted in WarpX mitigates both spurious effects. 

The mesh refinement implementation was first tested on simple test cases (single particle dynamics with self-field, synchrotron radiation emission, beam oscillations), and then on the modeling of BPA and LPA accelerators, as described below.
 
\section{Application to the modeling of plasma-based accelerators with mesh refinement.}

This subsection presents results demonstrating the first WarpX simulations with mesh refinement of particle beam-driven plasma accelerator (BPA) and laser-driven plasma accelerator (LPA), in 2-D and 3-D. The main physical and numerical parameters are summarized in Table \ref{Table_pLPA}.

\begin{table*}[htp]
\footnotesize
\caption{Main physical and numerical parameters of the WarpX simulations of BPA and LPA plasma accelerators.}
\begin{center}
\begin{tabular}{ll}

\hline
{\bf Plasma} & \\
\hspace{0.5cm}Plasma density & $10^{25}$ m$^{-3}$ \\

\hline
{\bf Witness electron beam} &\\
\hspace{0.5cm}Witness beam charge (BPA/LPA)& $-10$/$-20$ pC \\
\hspace{0.5cm}Witness beam profile & Gaussian \\
\hspace{0.5cm}Witness beam X R.M.S. size & 0.1 $\mu$m \\
\hspace{0.5cm}Witness beam Y R.M.S. size & 0.1 $\mu$m \\
\hspace{0.5cm}Witness beam Z R.M.S. size & 0.2 $\mu$m \\
\hspace{0.5cm}Witness beam initial energy & $10$ MeV\\

\hline
{\bf Driver electron beam} &\\
\hspace{0.5cm}Driver beam charge & $-100$ pC \\
\hspace{0.5cm}Driver beam profile & Gaussian \\
\hspace{0.5cm}Driver beam X R.M.S. size & 3 $\mu$m \\
\hspace{0.5cm}Driver beam Y R.M.S. size & 3 $\mu$m \\
\hspace{0.5cm}Driver beam Z R.M.S. size & 0.2 $\mu$m \\
\hspace{0.5cm}Driver beam initial energy & $500$ GeV\\

\hline
{\bf Driver laser beam} & \\
\hspace{0.5cm}Laser beam max field & $14$ TV \\
\hspace{0.5cm}Laser beam profile & Gaussian \\
\hspace{0.5cm}Laser beam waist & 10 $\mu$m \\
\hspace{0.5cm}Laser beam duration & 5 fs \\
\hspace{0.5cm}Laser wavelength & 0.8 $\mu$m \\

\hline
{\bf Numerical parameters} & \\
\hspace{0.5cm}Spatial base \# grid cells BPA & $32 \times 32 \times 192$ \\
\hspace{0.5cm}Spatial base \# grid cells LPA & $32 \times 32 \times 512$ \\
\hspace{0.5cm}Simulation box size BPA & $30 \times 30 \times 15$ $\mu$m$^3$\\
\hspace{0.5cm}Simulation box size LPA & $40 \times 40 \times 20$ $\mu$m$^3$\\
\hspace{0.5cm}Averaged \# macroparticles/cell in plasma in main grid& $2 \times 2 \times 2$ \\
\hspace{0.5cm}Averaged \# macroparticles/cell in plasma in refinement patch& $1 \times 1 \times 1$ \\
\hspace{0.5cm}\# macroparticles in witness electron beam& $10000$ \\
\hspace{0.5cm}\# macroparticles in driver electron beam& $10000$ \\
\hspace{0.5cm}Particle shape factor & Cubic \\

\hline
\end{tabular}
\end{center}
\label{Table_pLPA}
\end{table*}%

In all the simulations presented in this section, the density of the witness electron beam is high enough to generate self-fields with magnitudes that are comparable to the fields generated by the wake of the driver particle or laser beam. Since the dimensions of the witness beam are much smaller than the ones of the wake, the grid resolution that is needed to capture those self-fields accurately is higher than what is needed to resolve the wake. Hence mesh refinement can be used effectively to increase the resolution in the volume immediately surrounding the witness beam.


\begin{figure}[!tbh]
  \centering
  \includegraphics[width=8.cm]{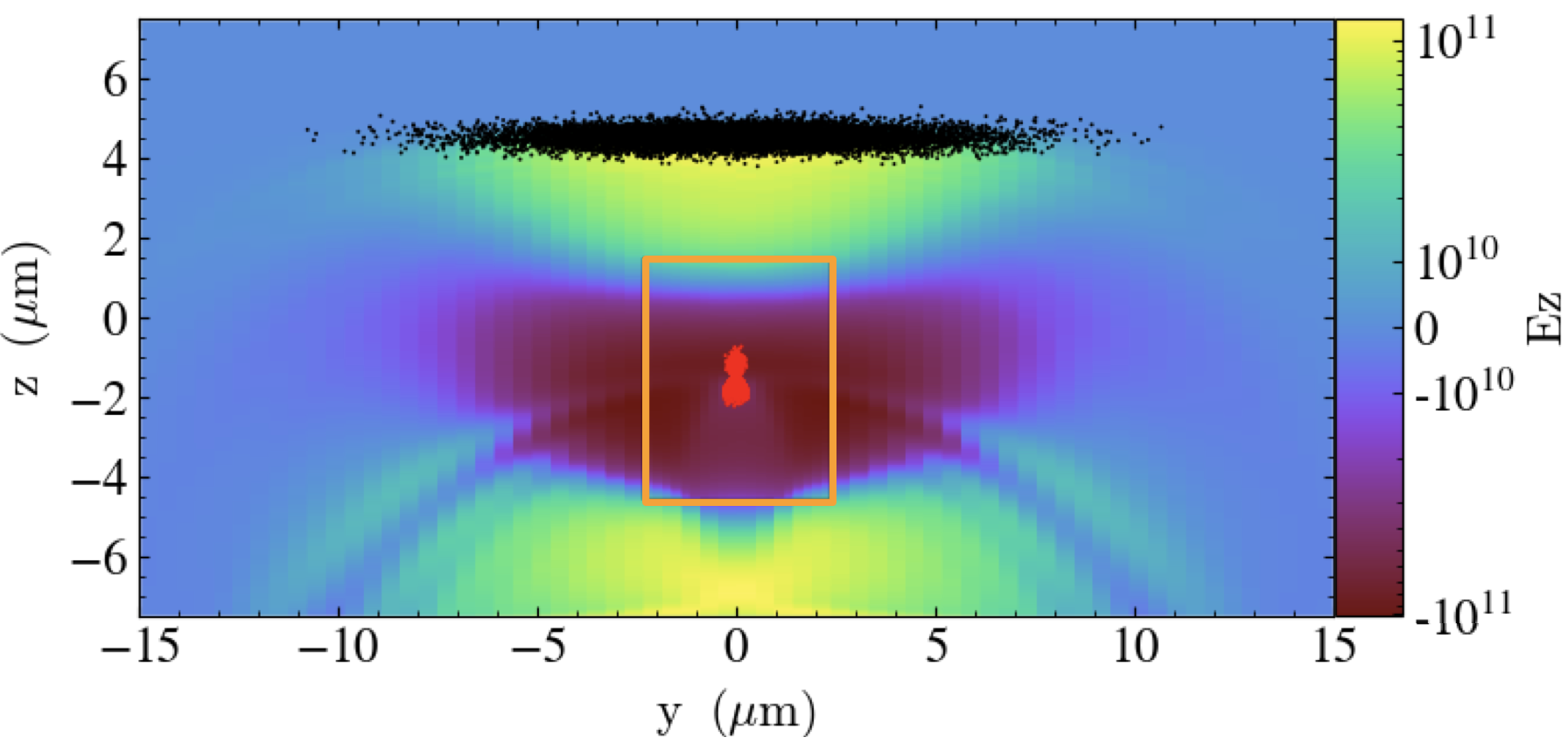}
  \includegraphics[width=9.cm]{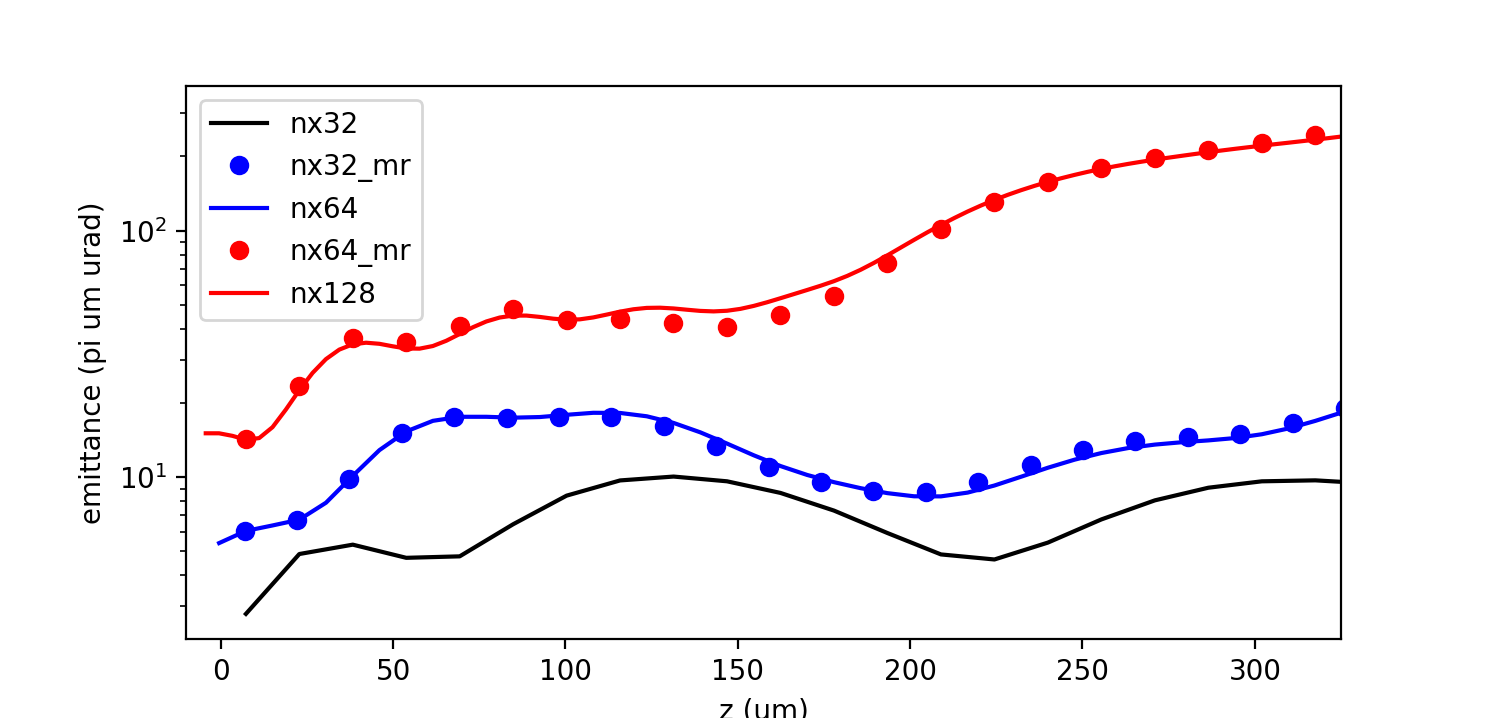}
  \caption{Results from the BPA simulations in 3D showing: (top) snapshot of the longitudinal electric field map, with overlaid beam particle drive (black dots), witness electron beam (red dots), and limits of the mesh refinement patch (orange lines); (bottom) evolution of the witness electron beam emittance as a function of propagation distance for simulations at various resolutions with and without mesh refinement (curves are labeled according to the number of transverse grid points).}
    \label{fig:BPA_3D}
\end{figure}

\begin{figure}[!tbh]
  \centering
  \includegraphics[width=8.cm]{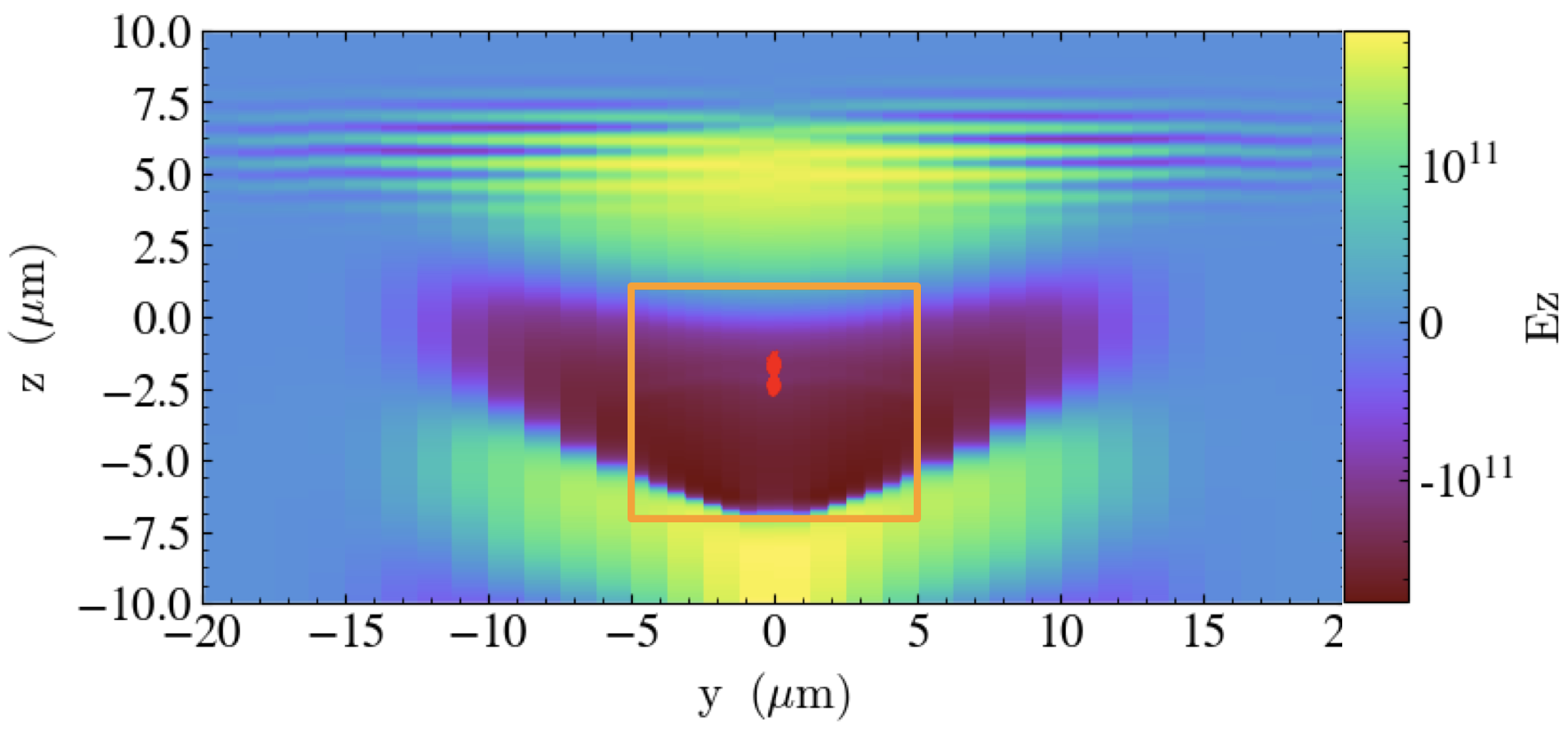}
  \includegraphics[width=8.cm]{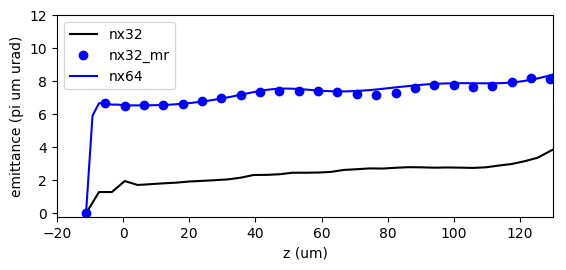}
  \caption{Results from the LPA simulations in 3D showing: (top) snapshot of the longitudinal electric field map, with overlaid witness electron beam (red dots), and limits of the mesh refinement patch (orange lines); (bottom) evolution of the witness electron beam emittance as a function of propagation distance for simulations at various resolutions with and without mesh refinement (curves are labeled according to the number of transverse grid points).}
    \label{fig:LPA_3D}
\end{figure}

Simulations of BPA and LPA plasma accelerators were performed in 2-D and 3-D, with a base spatial grid resolution that was increased progressively by a factor of 2 along each dimension, with and without mesh refinement. For the cases with mesh refinement, the resolution is twice the one of the parent grid along each dimension. The time step is set at the Courant limit of the smallest grid cell in the simulation.

Results from the 3-D simulations are presented in Fig. \ref{fig:BPA_3D} and Fig. \ref{fig:LPA_3D}. Simulations without mesh refinement show that, as expected, the witness electron beam emittance (a measure of beam quality, the lower the better for a particle collider) increases with resolution, as the beam field, which is much more localized than the wake from the driver beam (laser or electrons), goes from under-resolved to increasingly well resolved. On the other hand, in the quasi-linear regime that was chosen here, the wake from the driver beam is fairly smooth and should be fairly well resolved, even at the lowest resolution. Simulations with a mesh refinement patch surrounding the witness beam (at twice the resolution of the parent grid along each dimension) can reproduce the results of simulations without mesh refinement at twice the resolution, reinforcing that: (a) the mesh refinement patch is very effective at increasing the resolution locally for the very localized witness electron beam self-fields, and (b) the smooth wake generated by the driven beam seems well resolved at moderate resolution, since relaxing the resolution everywhere except around the witness beam does leave the result practically unchanged.

\section{Plans}
The code continues to be developed, with additional Maxwell field solvers, such as ultrahigh-order pseudo-spectral analytical time-domain (PSATD) \cite{VayJCP2013, VincentiCPC2016, Blaclard2017, Leblanc2017, Vincenti2017a}, multiple levels of mesh refinement, tunnel and impact ionization, etc, as required for the modeling of plasma accelerators and other possible applications. The code will be open source and made available to the public toward the end of 2018.

\section{ACKNOWLEDGMENT}
This work was supported by the Exascale Computing Project (17-SC-20-SC), a collaborative effort of two U.S. Department of Energy organizations (Office of Science and the National Nuclear Security Administration) responsible for the planning and preparation of a capable exascale ecosystem, including software, applications, hardware, advanced system engineering, and early testbed platforms, in support of the nation's exascale computing imperative. Used resources of the National Energy Research Scientific Computing Center.

This document was prepared as an account of work sponsored in part
by the United States Government. While this document is believed to
contain correct information, neither the United States Government
nor any agency thereof, nor The Regents of the University of California,
nor any of their employees, nor the authors makes any warranty, express
or implied, or assumes any legal responsibility for the accuracy,
completeness, or usefulness of any information, apparatus, product,
or process disclosed, or represents that its use would not infringe
privately owned rights. Reference herein to any specific commercial
product, process, or service by its trade name, trademark, manufacturer,
or otherwise, does not necessarily constitute or imply its endorsement,
recommendation, or favoring by the United States Government or any
agency thereof, or The Regents of the University of California. The
views and opinions of authors expressed herein do not necessarily
state or reflect those of the United States Government or any agency
thereof or The Regents of the University of California. 


\bibliographystyle{elsarticle-num}
\bibliography{EAAC2017_WG6_Vay}




\end{document}